# Accurate quantification of protein concentrations and state populations by 2D $^{1}H$-$^{15}N$ Q-BEST NMR

Yacine Ouali, Beate Bersch, and Bernhard Brutscher *

Université Grenoble Alpes, CEA, CNRS, Institut de Biologie Structurale (IBS), 71 avenue des Martyrs, 38044 Grenoble Cedex 9, France

* To whom correspondence should be addressed:

E-mail: bernhard.brutscher@ibs.fr
Phone: (33) 4 57 42 85 62
ORCID: 0000-0001-7652-7384

### *Abstract*

We introduce 2D Q-BEST NMR as a quantitative approach for determining protein concentrations and relative state populations from a set of three $^{1}H$–$^{15}N$ correlation spectra. The BEST methodology ensures short overall experimental times and highest sensitivity. The method was validated using ubiquitin as a model system, where a precision of approximately 15–20% was achieved at the level of individual cross-peaks. This precision can be further improved by averaging corrected intensities of well-resolved resonances corresponding to the same protein state. The approach was subsequently applied to the quantification of protein concentrations across multiple samples and to the determination of conformational state populations in proteins exhibiting intrinsic or ligand-induced heterogeneity. In both cases, 2D Q-BEST NMR provides accurate and robust quantification of NMR-visible species, demonstrating its utility as a general spectroscopic tool for quantitative biomolecular NMR analysis.

## *Introduction*

Quantitative nuclear magnetic resonance (qNMR) spectroscopy is a powerful analytical technique that enables the direct determination of relative concentrations of chemical species in solution by comparing NMR signal intensities detected for different chemical species. Absolute concentrations can be obtained by performing measurements in the presence of an internal or external standard of known purity and concentration, providing a direct reference for intensity calibration. qNMR has found widespread use in various research and application fields [1–3]. For example, in pharmaceutical analysis, it is applied for purity assessment and content determination [4]. In metabolomics, qNMR enables reproducible and non-targeted quantification of metabolites in complex biological mixtures [5], while in food chemistry it provides information on product authenticity and adulteration [6,7].

One-dimensional qNMR (1D qNMR) exploits the linear relationship between the integrated NMR signal intensity and the number of observed nuclei. Following a simple single-pulse excitation, the detected NMR signal is proportional to the Zeeman spin population and, consequently, to the number of nuclear spins within the NMR-active sample volume. Accurate quantification requires the following experimental conditions to be satisfied: (i) uniform excitation of all spins, and (ii) the use of an inter-scan delay ($T_{rec}$) sufficiently long to allow full longitudinal relaxation. In practice, $T_{rec}$ should be at least five times the longitudinal relaxation time ($T_1$) of the nuclei under observation ($T_{rec} > 5*T_1$), ensuring that the deviation of steady state spin population from thermal equilibrium is less than 1% .

The situation is more complex in 2D NMR, requiring a multi-pulse sequence that may lead to differential signal decay of individual nuclear spins due to differences in nuclear spin interactions and molecular dynamics that govern spin relaxation, thus breaking the linear relationship between NMR signal volumes and relative molecular concentrations. The practical aspects of quantitative 2D NMR and its applications to the analysis of complex mixtures have recently been comprehensively reviewed by Giraudeau and coworkers[8]. A particularly elegant solution to 2D qNMR was proposed by the Markley group [9] in the context of 2D $^{1}H$-$^{13}C$ correlation NMR spectroscopy of metabolite mixtures. Their approach, termed "extrapolated time-zero HSQC ($HSQC_0$)", effectively removes pulse-sequence-induced differential signal attenuation of individual nuclear spins (chemical species). This strategy restores the proportionality between signal volumes and molecular concentrations, thereby providing a reliable basis for quantitative analysis. The $HSQC_0$ approach has recently also been applied to determine the partitioning of metabolites into biomolecular condensates [10].

In the $HSQC_0$ approach, the pulse sequence is repeated several (N) times, enabling back-extrapolation of the NMR signal intensities (volumes) to the hypothetical case in which no pulses or delays, other than a single 90° excitation pulse, are applied. The extrapolated signal intensity is given by:

$$V^0 = \frac{V^N}{(k_{loss})^N} \quad \text{with} \quad k_{loss} = \frac{V^2}{V^1} = \frac{V^3}{V^2} \tag{1}$$

Here, $V^N$ denotes the NMR signal volume obtained after N repetitions of the fundamental pulse-sequence module, and $k_{loss}$ represents the attenuation factor per module. This factor accounts for cumulative signal losses arising from spin relaxation, pulse imperfections, homonuclear and heteronuclear scalar coupling evolution, as well as chemical exchange processes that affect the nuclear spins during each repetition of the specific pulse-sequence element (e.g., HSQC or TROSY building blocks).

In this contribution, we extend the $HSQC_0$ concept to 2D $^1$H-$^{15}$N experiments of proteins to provide an NMR-based method for (i) accurate concentration determination of protein NMR samples, and (ii) quantification of relative state populations under solution conditions where the protein adopts various conformational states that are in slow exchange on the NMR chemical shift time scale. In particular, we propose a sensitivity-optimized experimental protocol for 2D qNMR, based on BEST-HSQC [11,12] and BEST-TROSY [13,14] pulse schemes, and evaluate the performance of this 2D $^1$H-$^{15}$N Q-BEST NMR methodology for a selection of model protein systems.

## *Results and discussion*

### *2D Q-BEST NMR experiments*

Two distinct pulse schemes for 2D $^1$H-$^{15}$N Q-BEST NMR are shown in **figure 1**. The first one (**figure 1A**) is based on a sensitivity enhanced BEST-HSQC experiment, while the second one (**figure 1B**) exploits a BEST-TROSY pulse sequence. In both cases, the basic pulse-sequence module is repeated N times, yielding Q1-, Q2-, …, QN-type experiments. The (N−1) additional repetitions of the pulse sequence are modified as follows: (i) NMR signal acquisition, $t_1$ incrementation, and quadrature detection are performed only for the final pulse-sequence module; (ii) coherence-selection gradients are omitted in the additional (N−1) repetitions to preserve pure-phase coherence at the end of each module; and (iii) a final 90° $^{1}$H pulse (highlighted in grey) converts proton transverse single-quantum (SQ) magnetization into longitudinal spin polarization ($H_z$), thereby ensuring identical initial spin states at the beginning of each pulse-sequence module.

All $^{1}$H pulses are applied as band-selective shaped pulses, typically centered at 8.5 ppm and covering a band width of 5.0 ppm, which is the basis of the BEST (band-selective excitation short transient) concept that achieves longitudinal $^{1}$H relaxation enhancement (LRE) by leaving the water and aliphatic $^{1}$H at (or close to) thermal equilibrium [15,16]. This reduces the effective $^{1}$H relaxation time from about 1 s to 200 ms for amide protons in well-structured protein regions (α-helix, β-sheet), and from a few seconds to less than 100 ms for water-exposed, non-hydrogen-bonded amide protons (at neutral or basic pH) [15,17]. As a consequence, the optimal recycle delay ($T_{rec} \sim T_1$) required to achieve maximal sensitivity, defined as the signal-to-noise ratio per unit experimental time, is significantly reduced, while the overall sensitivity is increased by approximately 50%. An additional advantage in the context of quantitative NMR is the fact that Q-BEST sequences permit the use of comparatively short recycle delays, typically in the range of 1.0–2.0 s (as opposed to 5–10 s for conventional $^{1}$H-$^{15}$N qNMR experiments), while still satisfying the condition ($T_{rec} > 5*T_1$), which significantly improves the sensitivity of qNMR.

We first tested the 2D $^{1}$H-$^{15}$N Q-BEST NMR sequences on a 1 mM sample of $^{13}$C/$^{15}$N-labelled ubiquitin (8.6 kDa protein) at 15°C and 600 MHz $^{1}$H frequency. A series of Q1, Q2, and Q3 spectra were recorded setting the recycle delay to either $d_{rec}$ = 2s or $d_{rec}$ = 200 ms for both pulse schemes of figure 1. We also experimentally checked that the NMR peak intensities remained unchanged when further increasing the recycle delay from 2s to 5s, proving that the condition of 5*T1 is respected for this inter-scan delay. As pointed out in equation (1), $k_{loss}$ can be computed from a pair of experiments, either Q1 and Q2 or alternatively, Q2 and Q3 (and in principle also Q1 and Q3), recorded with the same inter-scan delay. **Figures 2A** and **2B** show correlation plots of residue-specific $k_{loss}$ values extracted from different experimental pairs. These data demonstrate that, as expected from theory, the same results are obtained whether the data were recorded with long or short recycle delays, and whether one (Q2) or two (Q3) sequence repetitions were used. The attenuation factor varies from 0.2 to 0.8 for individual amide groups in ubiquitin with an average of about 0.45, independent of the pulse sequence (Q-Best-HSQC or Q-Best-TROSY) used. This is in good qualitative agreement with signal attenuation estimations based on ubiquitin $^1$H and $^{15}$N relaxation times and pulse sequence delays.

In view of these findings and in order to optimize experimental sensitivity of the method, we propose a general protocol for quantitative 2D BEST NMR based on the recording of 3 separate data sets: (i) a first one with N = 1, and $T_{rec}$ > 5*T1 (Q1-ref); (ii) a second one with N = 1, and $T_{rec}$ optimized for maximal sensitivity (Q1-opt); (iii) a third one with N = 2, and $T_{rec}$ optimized for maximal sensitivity (Q2-opt). In addition, the number of sequence repetitions (scans) should be adjusted for each experiment separately to account for differences in NMR signal intensity, and thus allow for reliable intensity (volume) measurements in each data set. Corrected NMR peak volumes ($V^0$) are then be computed from these data as follows:

$$V^0 = \frac{V\,(Q1-ref)}{NS(Q1-ref)} \times k_{loss} \quad \text{with} \quad k_{loss} = \frac{I(Q1-opt)}{I(Q2-opt)} \times \frac{NS(Q2-opt)}{NS(Q1-opt)} \qquad (2)$$

While measurement of peak volumes is mandatory for the Q1-ref experiment, either peak volumes or peak intensities can be used to determine $k_{loss}$ from the Q1-opt and Q2-opt data sets. In this work, we chose to use peak intensities because they are generally less dependent on the peak-picking algorithm used.

The residue-specific NMR volumes measured for ubiquitin in the Q1-ref spectra span a range of a factor of 5, despite the fact that each resonance arises from the same number of nuclear spins, corresponding to the number of ubiquitin molecules within the NMR detection volume). This spread can be attributed to differences in the dynamics of the ubiquitin backbone on pico- to millisecond timescales, which result in variable relaxation-induced signal loss during the pulse sequence. When NMR intensities, rather than volumes, are extracted the variability becomes even more pronounced, with differences approaching an order of magnitude between the weakest and strongest signals (data not shown). These observations clearly demonstrate that, in standard 2D $^1$H-$^{15}$N NMR experiments, peak volumes (or intensities) do not provide a reliable basis for accurate determination of protein concentrations or conformational state populations. Applying the Q-BEST-NMR methodology, the spread in peak intensities ($V^0$) is reduced to about ± 15% for BEST-TROSY, and ± 25% for BEST-HSQC around the mean value. As the signal to noise ratio (SNR) of the different spectra recorded for ubiquitin is much higher than the observed variability of corrected peak intensities (SNR > 100), this indicates that ±15% may be an intrinsic limit in the achievable precision of 2D-based quantitative NMR measurements, most likely due to the difficulty of capturing the contribution of all spins to the NMR peak volume, as well as other spectral artefacts limiting the accuracy of peak volume measurement. The slightly better performance of BEST-TROSY compared to BEST-HSQC can be explained by the improved line widths of the NMR signals in both spectral dimensions as illustrated in **figure 2D** making volume measurements more reliable. In the present work, peak volumes were determined by Gaussian peak fitting using the implementation available in the CCPNMR-2 Analysis software [18]. The precision of 2D Q-BEST NMR may be further improved through the use of more advanced peak integration methods, such as the recently proposed deep learning-based DeepPicker algorithm [19].

*Protein concentration measurements*

A first application of 2D $^{1}H$-$^{15}N$ Q-BEST NMR is the determination of the concentration of the NMR-observable protein species in a given sample. Such an approach is particularly valuable in cases where conventional methods are unreliable, including: (i) proteins for which UV absorbance at 280 nm provides inaccurate estimates (e.g., proteins lacking aromatic residues, intrinsically disordered proteins (IDPs), or samples affected by background absorbance from buffer components); (ii) samples exhibiting partial aggregation, leading to the coexistence of NMR-visible and NMR-invisible species; and (iii) proteins with incomplete or only partial $^{15}N$ labelling. In principle, similar information could be obtained from a 1D $^{1}H$ NMR spectrum[20,21]. However, a 2D approach offers several important advantages. First, a larger number of individual resonances are spectrally resolved, improving statistical robustness. Second, spectral artefacts can be more readily identified and excluded, thereby reducing the risk of systematic quantification errors. Third, 2D $^{1}H$-$^{15}N$ experiments enable the direct and selective quantification of the isotope-labeled fraction of the protein.

For quantitative concentration analysis, a dedicated 2D Q-BEST template is required in which all acquisition and processing parameters are strictly fixed across the three experiments (Q1-ref, Q1-opt, and Q2-opt). All spectra must be recorded on the same NMR spectrometer under identical experimental conditions, including pulse-sequence parameters, receiver gain, coherence-transfer delays, acquisition times in both $t_1$ and $t_2$ dimensions, and spectral widths. Likewise, data processing must be performed in a fully consistent manner, using identical procedures for apodization, zero-filling, Fourier transformation, and baseline correction. The only parameter that may be varied between experiments is the number of scans, which is adjusted according to sensitivity requirements. The resulting differences are subsequently accounted for during intensity normalization, as described above (equation 2). This experimental strategy enables the determination of relative protein concentrations across different samples. For absolute quantification, however, a reference sample with a precisely known protein concentration is required. The corresponding dataset serves as external calibration standard, allowing the concentration of protein species in unknown samples to be determined by comparing the measured (or corrected) NMR intensities to those of the reference.

This approach is illustrated in **figure 3** for a set of uniformly $^{13}C/^{15}N$-labeled protein samples available in our laboratory, with molecular weights ranging from 8.6 to 26 kDa, and sample concentrations between 150 μM and 1.1 mM. One of the samples (Sir-FP18) was additionally ~70% deuterated. All datasets were acquired on a Bruker Avance IIIHD 600 MHz NMR spectrometer equipped with a cryogenically cooled TCI cryoprobe at 25°C, using sample geometries with either 5 mm or 3 mm tube diameter. 1 mM samples of Ubiquitin were used as reference. **Figures 3a** and **3b** present the results obtained for the 5mm and 3mm samples, respectively. The left panels show the peak volumes measured in the Q1-ref spectrum, whereas the right panels display the corrected peak volumes ($V^0$) calculated using equation (2). $^1H$-$^{15}N$ correlation peaks located in crowded spectral regions were excluded from the analysis, as were peaks with significantly higher than average intensity, potentially indicating peak overlap. In all cases, the quantitative 2D method yields concentration estimates deviating by less than 10% from the expected values, whereas uncorrected measurements can underestimate the protein concentration by more than a factor of three.

*Protein conformational state populations*

Proteins in solution populate dynamic ensembles of conformations that are often essential for their biological (or biotechnological) function. Such conformational dynamics span a wide range of timescales and amplitudes. In the case of slow dynamics (on the NMR chemical shift time scale) this results in conformational (or chemical) heterogeneity giving rise to distinct chemical shifts of nuclear spins in these different states (chemical environments). Such heterogeneous NMR spectra are also observed as a result of incomplete or heterogeneous complex formation in a binding assay, chemical reactions, protein degradation, or due to different sample environments, e.g. in-cell protein NMR observing proteins in different cellular compartments, or inside and outside the cell.

2D NMR offers a particularly powerful tool for studying protein dynamics and conformational heterogeneity in solution at atomic resolution. In particular, $^1H$-$^{15}N$ correlation spectra are highly sensitive to subtle structural rearrangements and chemical modifications, therefore providing residue-specific insights into different protein states. Accurate quantification of the populations of these states is essential for understanding the thermodynamics governing the observed conformational dynamics or binding equilibria.

Here, we applied the 2D Q-BEST method to two protein systems exhibiting heterogeneous $^{1}H$-$^{15}N$ correlation spectra. The first example is the GFP-like β-barrel protein mEos4b, a molecular label engineered for applications in super-resolution fluorescence microscopy [22,23]. mEos4b undergoes photoconversion from a green- to a red-fluorescent form upon UV illumination. Interestingly, in both fluorescent states (green and red) NMR experiments reveal conformational heterogeneity, with two coexisting conformations [24,25] that differ in their photophysical properties. In the green fluorescent form, the two conformational states, termed A and B, interchange on the minutes time scale at 35°C, and their relative populations can be modulated by pH or UV illumination [24]. To investigate this dynamic process in greater detail, accurate quantification of the A- and B-state populations from the NMR spectra is essential. **Figure 4A** shows the result of 2D $^{1}H$-$^{15}N$ Q-BEST measurements for the Green state of mEos4b at pH 7.5. The experimental data are presented as a correlation plot of the measured peak volumes in the Q1-ref spectrum versus the computed signal-attenuation factor $k_{loss}$, calculated from the peak intensities extracted from the Q1-opt and Q2-opt spectral pair. In this representation, the linear slope of the data points provides a measure of the relative state populations ($V^0_A$ and $V^0_B$) with $V(Q1\text{-ref}) = V^0 * k_{loss}$. Note that only amide sites with well-resolved A- and B-state correlation peaks were used for this data analysis. The population percentages (and standard deviations) derived from the uncorrected peak volumes V(Q1-ref) are 56 ± 8 % for the A-state and 44 ± 6 % for the B-state, while using the corrected $V^0$ values we obtain 58 ± 4 % for A and 42 ± 3 % for B. Thus, in this particular case, application of the 2D Q-BEST method improves the precision, but not the mean values, of the derived relative populations. This observation can be explained by the fact that the amide moieties used for the analysis exhibit similar motional properties, resulting in a comparable range of $k_{loss}$ values (i.e. signal loss during the pulse sequence).

The second example concerns the binding of a small molecule (i.e. organic fluorogen) to a protein scaffold (pFAST). FAST (fluorescence-activating absorption-shifting tag) is a hybrid chemogenetic system used for fluorescence microscopy that combines the versatility of small synthetic fluorophores with the selectivity of a genetically encoded protein label [26]. We recently discovered that some proposed fluorogens bind to the pFAST binding pocket in a heterogeneous manner, leading to multiple $^{1}H$-$^{15}N$ peaks for the same amide site in the NMR spectra. F**igure 4B** illustrates this peculiar behavior for the pFAST-HBR-3M pair where four distinct correlation peaks are observed for amides located near the binding site: one corresponding to the apo (ligand-free) protein (apo), one to a major bound conformation (M), and two additional peaks to minor bound states (m1 and m2). A 2D Q-BEST NMR analysis, again based on well-resolved pure-state $^{1}H$-$^{15}N$ correlation peaks, reveals that 47% of pFAST proteins are in the apo state, while 53% are in one of the three bound states. The partitioning between the various bound conformations is 55% (M) and 22.5% (m1 and m2). Unfortunately, the qNMR analysis does not permit discrimination between data points originating from the two minor bound states, suggesting that their populations are indistinguishable within the experimental uncertainty. In this application, we observe a significant difference between the relative state populations extracted from 2D NMR peak volumes and those obtained using the 2D Q-BEST method: the population of the two minor states inferred from peak volumes in a single 2D data set was 20%, while this value increased to 22.5% after $k_{loss}$ correction. This statistically significant discrepancy arises because the amide sites in the minor bound states (m1 and m2) experience greater signal loss during the pulse sequence (smaller $k_{loss}$) than those in the major bound state (M). Thus, for this particular application, the 2D Q-BEST method not only provides more accurate and precise population estimates, but the measured $k_{loss}$ values also indicate differences in local dynamics within the pFAST binding pocket between the minor and major states of the pFAST-HBR complex.

### *Conclusions*

In this work, we introduced 2D Q-BEST NMR for the accurate determination of protein concentrations and relative state populations from a series of three $^{1}H$-$^{15}N$ correlation spectra. The performance of this quantitative 2D NMR approach was first evaluated using the model protein ubiquitin, demonstrating that a precision of approximately 15–20% can be achieved at the level of individual cross-peaks. The factors currently limiting higher precision in single-peak measurements remain to be fully elucidated. In practice, however, the experimental precision can be improved by averaging corrected peak intensities across well-resolved resonances corresponding to the same protein state. The 2D Q-BEST NMR methodology was subsequently applied to the determination of protein concentrations across a series of samples, as well as to the quantification of relative populations in proteins exhibiting intrinsic or binding-induced conformational heterogeneity. In both cases, the results show that 2D Q-BEST NMR constitutes a robust and versatile spectroscopic approach for the quantification of NMR-active species in protein systems.

### *Acknowledgments*

The authors thank Isabel Ayala (IBS) for support with protein isotope labeling and production, as well as Yuriy Spinov and Ludovic Jullien for chemical synthesis of the HBR fluorogen. Financial support from the Agence Nationale de la Recherche (grant no. ANR-22-CE11-0011-01) is gratefully acknowledged. Y. Ouali acknowledges receipt of a PhD fellowship from CEA. This work used the NMR platform of the Grenoble Instruct-ERIC centre (ISBG; UAR 3518 CNRS-CEA-UGA-EMBL) within the Grenoble Partnership for Structural Biology (PSB), supported by FRISBI (ANR-10-INBS-0005-02) and GRAL, financed within the University Grenoble Alpes graduate school (Ecoles Universitaires de Recherche) CBH-EUR-GS (ANR-17-EURE-0003). IBS acknowledges integration into the Interdisciplinary Research Institute of Grenoble (IRIG, CEA).

***Figure captions***

**Figure 1**

2D $^{1}H$-$^{15}N$ Q-BEST NMR experiments based on (**A**) sensitivity-enhanced BEST-HSQC or (**B**) BEST-TROSY pulse schemes. The first block (in brackets) is only applied for N >1: Q2 (once) and Q3 (twice). Filled and open pulse symbols correspond to 90° and 180° rf pulses, respectively. Band-selective $^{1}H$ pulses are applied with EBURP-2 (E) time-reversed E-BURP2 (E*) and REBURP (R) pulse shapes [27], while broadband $^{1}H$ pulses (dashed bars) are applied with a BIP-720-20 pulse shape [28]. The transfer delays Δ and Δ ' are set to $1/(4J_{NH})$, with Δ' slightly shorter to reduce the intensity of residual anti-TROSY components [29]. In our standard data set, these delays are set to Δ= 2.6 ms for the initial INEPT transfer, Δ= 2.7 ms and Δ'= 2.5 ms for the back transfer. Phase cycling was performed as follows: $\varphi_1$ = x, -x, , $\varphi_2$ = x (experiment **A**), $\varphi_2$ = y (experiment **B**), $\varphi_3$ = y, and $\varphi_{rec}$ = x,-x for Q1. In Q2 and Q3, $\varphi_1^*$ and $\varphi_{rec}$ are cycled independently leading to 4-step and 8-step phase cycles, respectively. Echo/anti-echo quadrature detection in $t_1$ is achieved by the pulsed field gradients $G_1$, $G_2$ and $G_3$ with (**A**) $G_1/G_2 = \pm \gamma_H/\gamma N$, and inverting phase $\varphi_2$ for the anti-echo component, and (**B**) $(G_1 - G_2)/(G_3 - G_2) = \pm \gamma_H/\gamma_N$ and inverting phase $\varphi_2$ and $\varphi_3$ for the anti-echo component. The duration of these selection gradients was set to 1 ms, followed by a 200 µs recovery delay. The pulse sequences, and standard acquisition and processing parameters, are included in the NMRlib package [30] that is freely available at the following URL: https://www.ibs.fr/en/communication-outreach/scientific-output/software/nmrlib-2-0-ibs-pulse-sequence-tools-for-bruker-spectrometers?lang=en. The Bruker code of the pulse sequences is also provided as Supporting Information.

**Figure 2**

2D Q-BEST performance test. All data were recorded on a 1mM sample of uniformly $^{13}C$,$^{15}N$-labeled ubiquitin (8.6 kDa) at 15°C on a Bruker Avance IIIHD 600 MHz NMR spectrometer equipped with a cryogenically cooled 5mm TCI probe. Q1, Q2, and Q3 data sets were recorded with the Q-BEST-HSQC and Q-BEST-TROSY sequences of figure 1, and recycle delays $d_{rec}$ of either 2s or 200 ms. (**A**) $k_{loss}$ values computed from peak intensity measurements according to equation (2). The correlation of $k_{loss}$ values derived from either Q2/Q1 or Q3/Q2 intensity ratios is shown. Red data points were measured with $d_{rec}$= 2s, while black data points correspond to $d_{rec}$= 200ms; (**B**) Dependence of $k_{loss}$ values on the recycle delay $d_{rec}$. Red data points were calculated as Q2/Q1, while black data points correspond to Q3/Q2. (**C**) Corrected and uncorrected 2D NMR peak volumes plotted as a function of the ubiquitin sequence. Black data points correspond to the raw volumes measured ($V^1$) in the Q1-ref experiment with $d_{rec}$= 2s, while red data points show the corrected peak volumes ($V^0$) according to the Q-BEST method and equation (2). For each data set, the mean volume was calculated as the average over all quantified 2D peak volumes. The deviation (in %) of the measured peak volume with respect to the calculated mean volume is plotted as a function of the ubiquitin sequence. (**D**) Stack plot of a small spectral $^1H$-$^{15}N$ region illustrating the difference in line widths obtained with BEST-HSQC and BEST-TROSY. These data were recorded on a 26 kDa GFP protein at 35°C and 600 MHz.

**Figure 3**

NMR protein concentration measurements carried out on (**A**) 5-mm and (**B**) 3-mm samples at 25°C and 600 MHz. 2D Q-BEST NMR data sets of the following protein samples were measured for this analysis: ubiquitin (8.6 kDa, 1 mM, $^{13}C/^{15}N$-labeled), GFP1 (26 kDa, 0.2 mM, $^{13}C/^{15}N$-labeled), GFP2 (26 kDa, 1.1 mM, $^{13}C/^{15}N$-labeled), GFP3 (26 kDa, 0.15 mM, $^{13}C/^{15}N$-labeled), ColN (10.4 kDa, 0.3 mM, $^{13}C/^{15}N$-labeled), BamE (11.4 kDa, 0.26 mM, $^{13}C/^{15}N$-labeled), Sir-FP18 (18 kDa, 0.5 mM, $^{13}C/^{15}N/^{2}H$(70%)-labeled). Three data sets (Q1-ref, Q1-opt, and Q2-opt) were recorded for each sample. The number of scans of the individual data sets was adjusted to achieve sufficient signal-to-noise ratios for peak intensity (volume) quantification. The overall experimental time to record a full Q-BEST NMR data set ranged from 1 h – 20 h for the individual samples. The uncorrected peak volumes, extracted from Q1-ref, are plotted in the left panel graphs, while the corrected peak volumes, computed with equation (2), are plotted in the right panel graphs. Peak numbering is arbitrary. The average value for each dataset is indicated by a straight line, and the relative concentrations with respect to ubiquitin (1 mM) are given numerically.

**Figure 4**

2D Q-BEST NMR applied to two conformationally heterogeneous NMR samples. (A) the green form of the fluorescent protein mEos4b, and (B) the complex formed between an organic fluorogen (HBR-3M) and the pFAST protein. 2D Q-BEST-TROSY data were recorded on Bruker Avance IIIHD 600 or 700 MHz NMR spectrometers equipped with cryogenically cooled 5mm TCI probes at sample temperatures of 35°C (mEos4b) and 20°C (pFAST). The results are shown as a correlation plot between the measured peaks volumes ($V^1$) from the Q1-ref spectrum and the calculated signal loss factor $k_{loss}$. This representation highlights differences in the range of measured $k_{loss}$ values across conformational states, as observed for the major and minor bound states in panel (B). Data points corresponding to NMR peaks of different conformational states are color-coded in (A) and annotated in (B). Note that the labels ‘m1’ and m2’ have been assigned arbitrarily to the two minor peaks. The light background color highlights the spread of uncorrected peak volumes measured for the individual conformational states. The slope of a linear fit of these data ($V^0$) provides a measure of the relative state populations. In addition, an annotated small spectral region and a drawing of the protein system is shown on top of the experimental data graph.

**Figure 1**

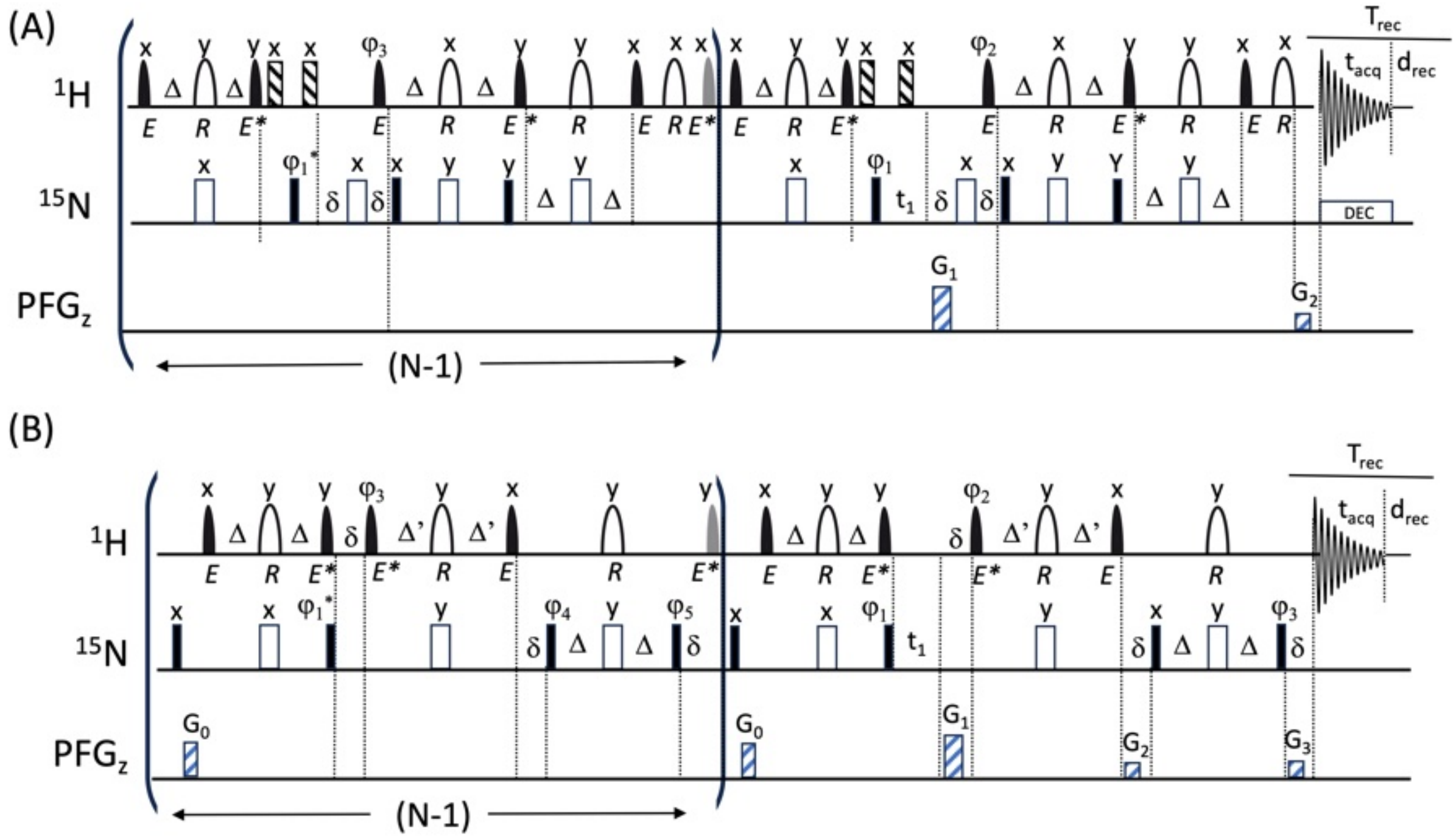

**Figure 2**

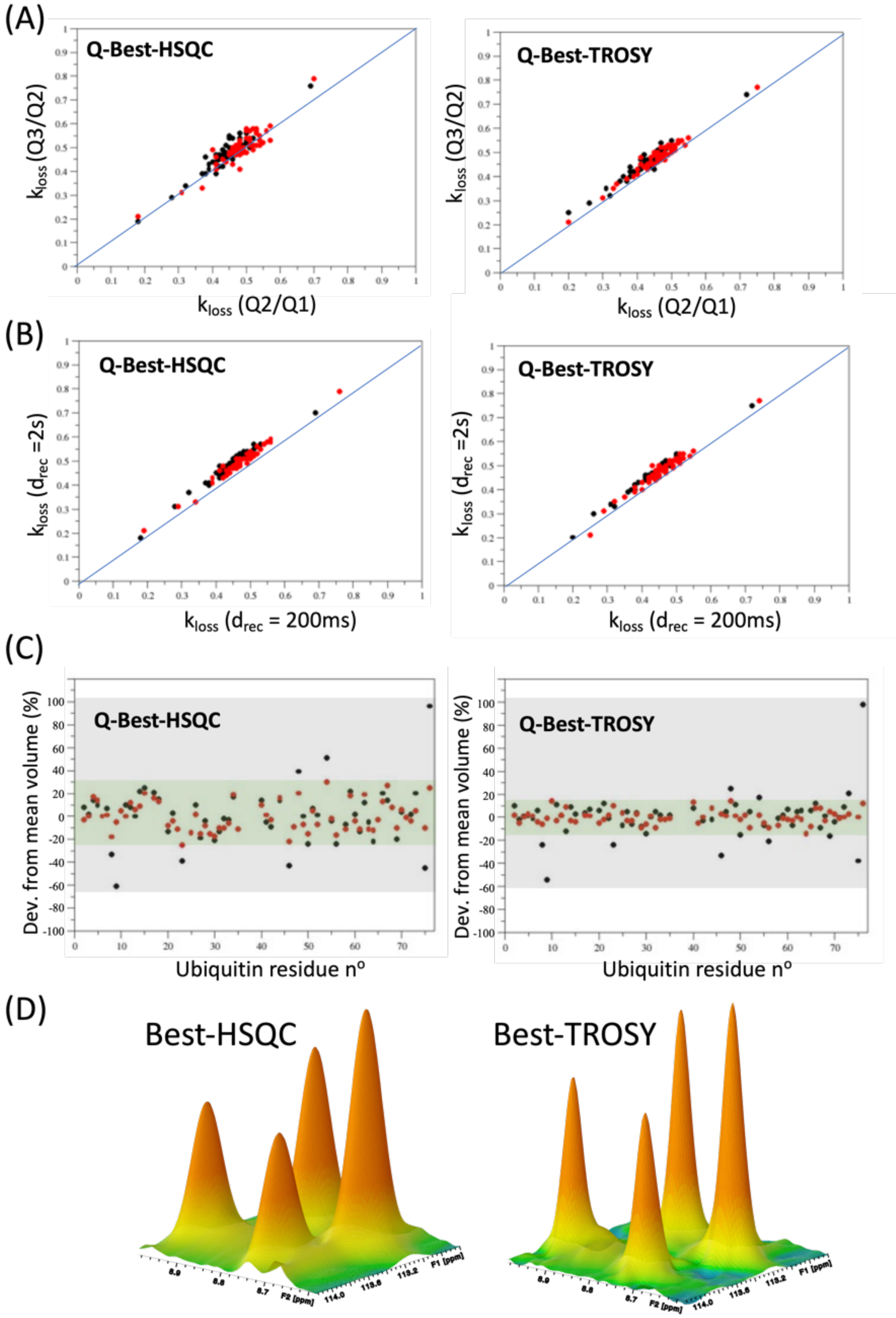

(A)
Q-Best-HSQC
$k_{loss}$ (Q3/Q2)
$k_{loss}$ (Q2/Q1)
Q-Best-TROSY
$k_{loss}$ (Q3/Q2)
$k_{loss}$ (Q2/Q1)
(B)
Q-Best-HSQC
$k_{loss}$ ($d_{rec}$ =2s)
$k_{loss}$ ($d_{rec}$ = 200ms)
Q-Best-TROSY
$k_{loss}$ ($d_{rec}$ =2s)
$k_{loss}$ ($d_{rec}$ = 200ms)
(C)
Q-Best-HSQC
Dev. from mean volume (%)
Ubiquitin residue n°
Q-Best-TROSY
Dev. from mean volume (%)
Ubiquitin residue n°
(D)
Best-HSQC
Best-TROSY

**Figure 3**

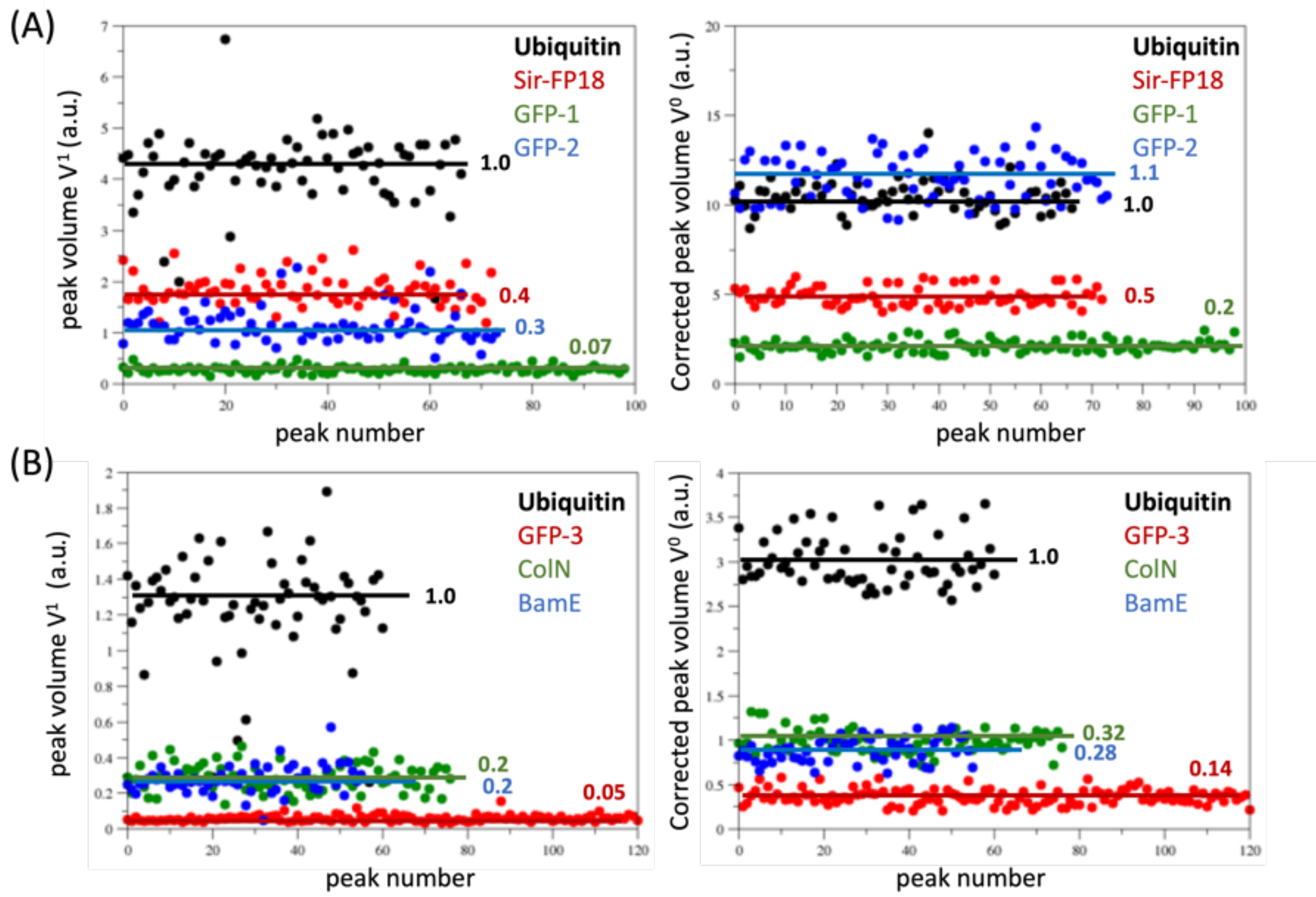

(A)
Ubiquitin
Sir-FP18
GFP-1
GFP-2
1.0
0.4
0.3
0.07
peak volume V¹ (a.u.)
peak number
Corrected peak volume V⁰ (a.u.)
1.1
1.0
0.5
0.2
(B)
Ubiquitin
GFP-3
ColN
BamE
1.0
0.2
0.2
0.05
0.32
0.28
0.14

**Figure 4**

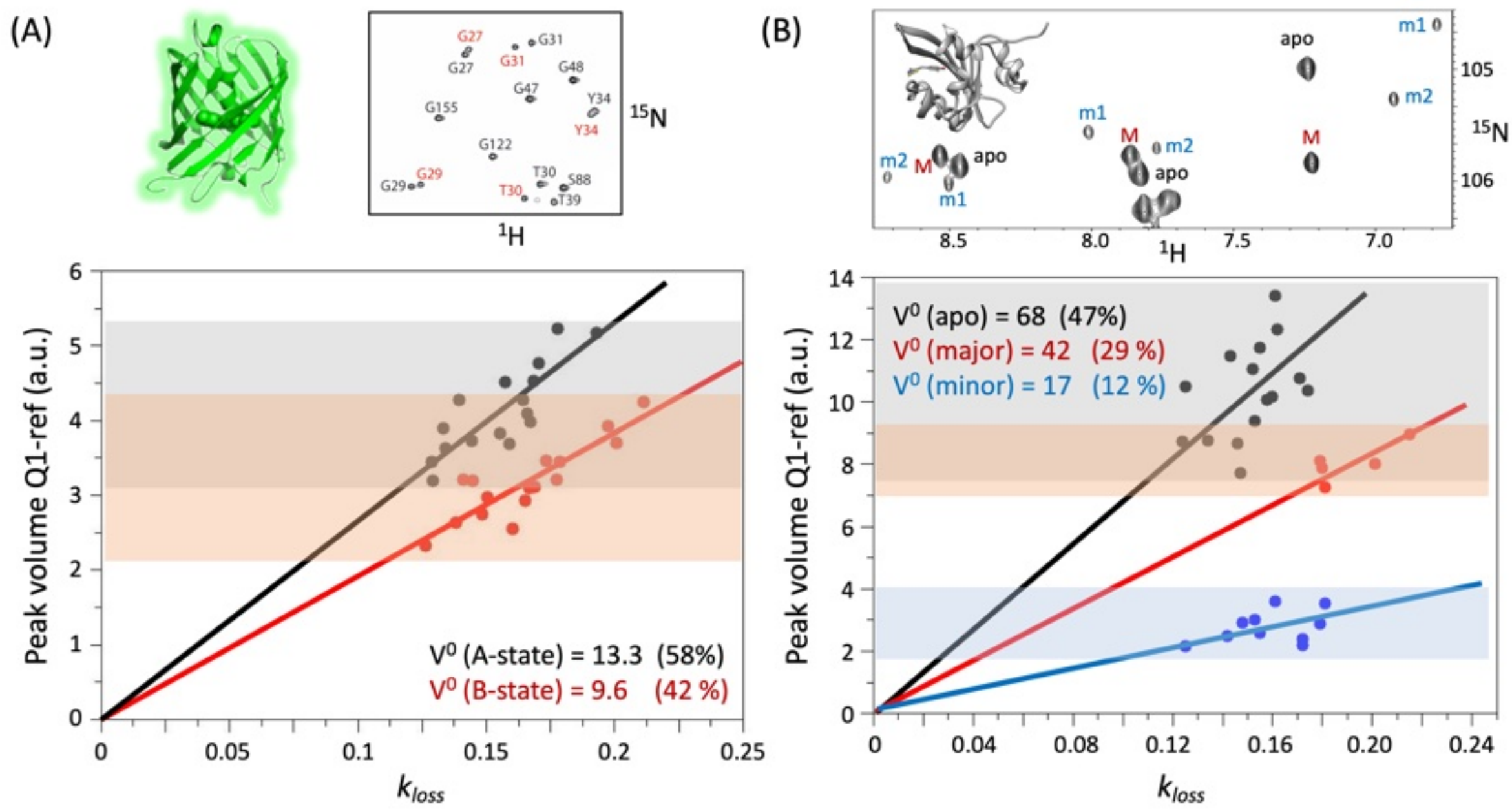

(A)
(B)
$V^0$ (A-state) = 13.3 (58%)
$V^0$ (B-state) = 9.6 (42 %)
$V^0$ (apo) = 68 (47%)
$V^0$ (major) = 42 (29 %)
$V^0$ (minor) = 17 (12 %)
Peak volume Q1-ref (a.u.)
$k_{loss}$
$^1$H
$^{15}$N